# Design Principles and Identification of Birefringent Materials

*Gwan Yeong Jung[1*], Guodong Ren[2], Pravan Omprakash[2], Jayakanth Ravichandran[3,4], Rohan Mishra[1,2*]*

[1]Department of Mechanical Engineering & Material Science, Washington University in St. Louis, St. Louis, MO 63130, USA

[2]Institute of Materials Science & Engineering, Washington University in St. Louis, St. Louis, MO 63130, USA

[3]Mork Family Department of Chemical Engineering and Materials Science, University of Southern California, Los Angeles, California 90089, USA

[4]Ming Hsieh Department of Electrical and Computer Engineering, University of Southern California, Los Angeles, California 90089, USA

*Corresponding author. Email: gjung@wustl.edu, rmishra@wustl.edu

**ABSTRACT**

Birefringence ($\Delta n$) is the dependence of the refractive index of a material on the polarization of light travelling through it. Birefringent materials are used as polarizers, waveplates, and for novel light-matter coupling. While several birefringent materials exist, only a handful of them show large $\Delta n > 0.3$, and are primarily limited to the infrared region. The variation of $\Delta n$ across diverse materials classes and strategies to achieve highly birefringent materials with transparency covering different regions of the electromagnetic spectrum are missing. We have calculated the $\Delta n$ of 967 non-cubic, formable crystals having vastly different structures, polyhedral connectivity and chemical compositions. From this set of compounds, we have screened highly-birefringent crystals ($\Delta n > 0.3$) having transparency in different regions of the electromagnetic spectrum. The screened compounds belong to several families such as $A'_3 M N_3$, $AMO_2$, $AN_3$, and $A'N_6$ ($A$ = Li, Na, K; $A'$ = Ca, Sr, Ba; $M$ = V, Nb, Ta). By analyzing the electronic structures of these compounds, we have distilled rules to enable the design of crystals with large $\Delta n$.

# 1. INTRODUCTION

In an anisotropic material, the complex index of refraction, $(n + i\kappa)$, is different along different principal axes of the crystal. Optical anisotropy is characterized by birefringence and dichroism, which are, respectively, the differences in the real parts ($\Delta n$) and the imaginary parts ($\Delta \kappa$) of the indices along different crystallographic orientations. Materials with large birefringence are critical in modern optics and are used in devices for optical communication,[1] laser-based technologies,[2, 3] imaging,[4] polarization control[5, 6] and wavefront manipulation.[7] Birefringent materials having transparency in different wavelengths of the electromagnetic spectrum, ranging from ultraviolet (UV) to infrared (IR), can enable the miniaturization of optical devices for aforementioned applications.[8-10] Commercial birefringent crystals, such as calcite, $YVO_4$,[11] $LiNbO_3$,[12] and α-$BaB_2O_4$,[13] have relatively low $\Delta n < 0.3$ with transparency in the UV to visible region. Moreover, these crystals face several limitations that hinder practical applications, including difficulties in growing high-quality single-crystals, relatively high costs, and constraints in thermal, mechanical, or chemical stability.[14] Recently, giant optical anisotropy with $\Delta n = [0.7, 2.1]$ has been reported in two-dimensional (2D) and quasi-one-dimensional (quasi-1D) materials such as $MoS_2$,[15] $As_2S_3$,[16] $BaTiS_3$,[17] and $Sr_{9/8}TiS_3$;[18] however, the transparency of these low-dimensional materials is restricted to the mid-IR region. Thus, there is an open opportunity to discover highly birefringent materials, which we define as $\Delta n > 0.3$, with transparency at operative working wavelengths ranging from IR to UV.

To achieve a large birefringence, several mechanisms have been proposed. These include: 1) introducing π-conjugated electrons through anions such as carbonates[19] and borates;[20] 2) incorporating highly polarizable ions[21] such as cations having lone-pair electrons like $Sn^{2+}$ or electronegative anions such as $F^-$; and 3) utilizing $d^0$ cations that are susceptible to second-order

Jahn-Teller distortions.[22] These mechanisms rely on polarizing, primarily, *s*- or *p*-electronic states near the Fermi energy.[23] However, as we show later, we find that highly birefringent band insulators, wherein electronic transitions occur from filled *s*- or *p*-like states to the empty conduction band, typically, have large band gaps that constrain transparency to the visible or UV ranges. In seeking birefringent materials for the IR region, alternative mechanisms have been explored. For example, Mei *et al.*[18] reported colossal birefringence ($\Delta n$ ~2.1) in the mid-IR region in a quasi-1D sulfide, $Sr_{9/8}TiS_3$. They attributed the origin of the colossal birefringence to the presence of highly oriented valence electrons residing in the $d_{z^2}$-states of specific Ti atoms — that can preferentially couple to light polarized along the long axis of $Sr_{9/8}TiS_3$. This mechanism, in contrast to the conventional mechanisms relying on anion *s*/*p*-electronic states, underscores how altering the character of the valence electrons near the Fermi energy can enable large $\Delta n$ – through directional, covalent *d-d* interactions between cations, rather than through anion-driven polarizability. To further advance the field and identify highly birefringent materials for targeted wavelengths, design strategies and exploration of a broad range of materials classes are needed.

In this Article, we have identified highly birefringent crystals ($\Delta n > 0.3$) with transparency spanning the UV to IR range using high-throughput density-functional theory (DFT) calculations. We computed the $\Delta n$ of 967 non-cubic crystals from the Materials Project database that are within their respective formability limits[24] and include only one anion.[25] These crystals span six materials classes including oxides, nitrides, sulfides, chlorides, fluorides and phosphides. Among the screened compounds, several families, such as $A'_3MN_3$, $AMO_2$, $AN_3$, and $A'N_6$ ($A$ = Li, Na, K; $A'$ = Ca, Sr, Ba; $M$ = V, Nb, Ta), are found to be highly birefringent ($\Delta n > 0.3$). By analyzing the electronic structures of these compounds, we outline design rules for large optical anisotropy and survey birefringent crystals that maintain transparency from UV to IR. Our study identifies three key strategies to achieve large $\Delta n$: 1) modulating the electron filling of the topmost valence band, 2) adjusting the anisotropy of the band dispersion near the Fermi energy, and 3) optimizing the spatial density and arrangement of polarizable anions such as $N_3^-$. Collectively, these findings will provide a framework for exploring highly birefringent materials across a vast structural and chemical space to eventually enable efficient light manipulation in nanophotonic and optoelectronic devices.

## 2. RESULTS & DISCUSSION

**2.1. Evolution of birefringence with varying structural anisotropy and electron filling in the prototypical perovskite $BaTiO_3$.** Motivated by the recent reports of giant birefringence in $BaTiS_3$[26] and $Sr_{9/8}TiS_3$,[18] both derivatives of the perovskite framework, we begin by investigating the role of structural anisotropy and electron filling on the optical anisotropy of the prototypical perovskite, barium titanate. We considered three phases of $BaTiO_3$ having different arrangements of the $TiO_6$ octahedra and varying structural anisotropy: 1) $BaTiO_3$-$R3m$ having isotropically arranged $TiO_6$ octahedra with three-dimensional (3D) corner connectivity, as shown in **Fig. 1a**[27]; 2) $BaTiO_3$-$Cmc2_1$ having a quasi-1D structure with chains of face-shared $TiO_6$ octahedra, as shown in **Fig. 1b**; and 3) $Ba_{9/8}TiO_3$-$R3c$, which displays a non-stoichiometric, modulated structure, as shown in **Fig. 1c**.[28] In this third structure, excess Ba atoms are accommodated in the lattice via a periodic rearrangement of the $TiO_6$ network. Modulation along the $c$-axis results in quasi-1D chains that alternate between segments of face-sharing octahedral (referred to 'O', colored in blue) and distorted trigonal prismatic units (referred to 'T', colored in orange). In these chains, the O and T units leads form a T-O-T-block that is followed by a block of five face-sharing O units. This

sequence is repeated twice, [-(T-O-T)-(O)$_5$-]$_2$ resulting in a periodicity corresponding to 16 TiO$_6$ units interleaved within 18 Ba layers.[18] A convex-hull analysis based on the formation enthalpy of the decomposition products indicates that only the *R*3*m* phase is on the hull, while *Cmc*2$_1$ and *R*3*c* phases are metastable being 127 and 53 meV/atom above the hull, respectively.

We track the electronic structure changes in these three polymorphs by focusing on the occupied electronic states near the Fermi energy, as these electrons at the top of the valence band are most susceptible to be polarized by the electric field of the incident light, and contribute to the real part of the index of refraction.[29] We calculated the electronic band structures of BaTiO$_3$, in *R*3*m* and *Cmc*2$_1$ phases and Ba$_{9/8}$TiO$_3$ in the *R*3*c* phase (**Fig. S1**). In both the *R*3*m* and *Cmc*2$_1$ phases of BaTiO$_3$, the valence bands near the Fermi energy are primarily composed of oxygen 2*p* states (**Fig. S1a,b** and **Fig. 1a,b**). Despite their similar electronic features, the $\Delta n$ of the *Cmc*2$_1$ phase is 0.3 compared to 0.1 for the *R*3*m* phase, as shown in **Fig. 1d**. We attribute this enhancement in $\Delta n$ to the increased structural anisotropy of the *Cmc*2$_1$ phase due to the quasi-1D chains of face-sharing TiO$_6$ octahedra. In contrast, the non-stoichiometric Ba$_{9/8}$TiO$_3$-*R*3*c* phase includes additional electrons from excess Ba$^{2+}$ ions that occupy the nominally empty Ti 3*d* states (**Fig. S1c**). This occupation changes the electronic character at the top of the valence band – from dispersed oxygen *p*-states in stoichiometric BaTiO$_3$ (colored as purple in *R*3*m*, and *Cmc*2$_1$) to localized $d_{z^2}$-states originating from the crystal-field splitting of Ti atoms in Ba$_{9/8}$TiO$_3$ (colored as green in *R*3*c*). Furthermore, a real-space analysis of the charge density shows that the electrons are localized on specific Ti atoms and are oriented along the *c*-axis. This shift in the character of the valence band aligns with previous findings reported in Sr$_{9/8}$TiS$_3$[18]. Comparing *Cmc*2$_1$-BaTiO$_3$ and *R*3*c*-Ba$_{9/8}$TiO$_3$, both of which have quasi-1D chains but have different electronic states that get polarized due to an incident light, we observe a further increase in $\Delta n$ from 0.3 to 1.0 in $R3c$-Ba$_{9/8}$TiO$_3$, as shown in **Fig. 1d**.

These changes in the electronic character also change the nature of the electronic transitions, which in turn shift the material's optical absorption edges, and its dichroic properties. In stoichiometric BaTiO$_3$-$R3m$ and BaTiO$_3$-$Cmc2_1$, the electronic transitions are of $p$-$d$ type that is typical in a band insulator: from the valence bands having primarily oxygen $p$ character to the empty conduction bands made up of Ti $3d$ states. Using the Perdew-Burke-Ernzerhof (PBE) functional with Hubbard $U$ correction ($U$ = 3.0 eV for Ti), the theoretical band gap of BaTiO$_3$-$R3m$ and BaTiO$_3$-$Cmc2_1$ is calculated to be 2.82 and 2.43 eV, respectively. A survey of birefringent band insulators (as discussed in section 2.2) shows that they have band gaps exceeding 2 eV, which makes them suitable for use in visible to UV regions. In contrast, $R3c$-Ba$_{9/8}$TiO$_3$, with its partially filled Ti $d_{z^2}$-states, exhibits a $d$-$d$ transition that is typical of a Mott insulator,[30] and has a reduced bandgap of 0.75 eV. This shifts the working wavelength region to the infrared. We note that $R3c$-Ba$_{9/8}$TiO$_3$ has a smaller $\Delta n$ = 1.0 compared to Sr$_{9/8}$TiS$_3$ ($\Delta n \approx 2.1$), despite having the same structural motif and Ti$d_{z^2}$-states dominating the top of the valence band. We attributed this difference to the more ionic character of Ti–O bonds compared to Ti–S bonds that are more covalent.[31] Overall, these findings hint that by changing the spatial arrangement of the polyhedral units and the electronic character of the top valence band, one can obtain birefringent compounds optimized for different wavelengths.

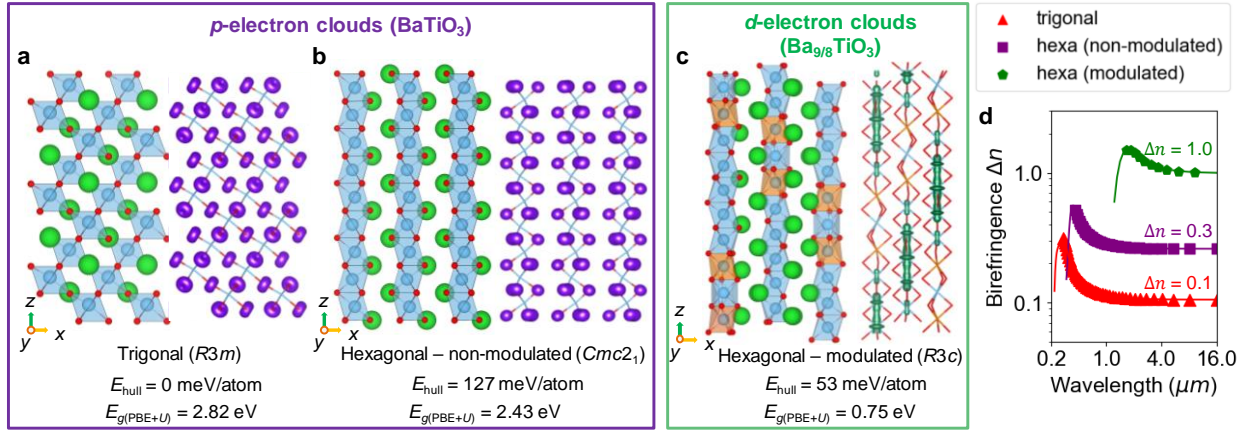

**Figure 1**. Crystal structure, electronic structure, and optical properties of two polymorphs of $BaTiO_3$ and $Ba_{9/8}TiO_3$. (a-c) Crystal structures and corresponding charge density isosurfaces are shown: (a) $BaTiO_3$ in $R3m$ phase, (b) $BaTiO_3$ in $Cmc2_1$ phase, and (c) $Ba_{9/8}TiO_3$ in $R3c$ phase. Spatial distribution of the valence electrons within 0.5 eV of the Fermi energy, showing O-$2p$ character in (a,b) and Ti-$\mathbf{3d_{z^2}}$ character in (c). The isosurface is set to an electron density of 0.005 e/Å³. (d) Calculated $\mathbf{\Delta n}$ of the three compounds.

**2.2. Key factors for large birefringence.** We next extended our analysis to encompass more structures from various materials classes, and identify birefringent compounds suitable for different wavelengths. 2D van der Waals layered structures often exhibit large $\Delta n$ due to the strong in-plane covalent bonds and weak out-of-plane van der Waals bonds.[15,16] In such 2D materials, the optic axis is typically oriented perpendicular to the layers, which implies that the large anisotropy in the plane is less accessible for practical device integration.[6, 18] Therefore, we narrowed down the scope of our search to bulk crystals that are more suitable for integration into optical devices. We considered six classes of materials based on anions with varying electronegativity – oxides, nitrides, fluorides, phosphides, sulfides and chlorides (**Fig. 2a**). The difference in the electronegativity of the cations and anions influences the electronic interactions and the band gap. We further classified the compounds based on the coordination geometry of the polyhedral building units, which determines the crystal field around the cations and its electronic structure (**Fig. 2b**, and **Fig. S2**). We labelled the groups using a combination of their polyhedral shape — that is the geometry of the surrounding anions around a central cation (e.g., linear, trigonal planar, square planar, and octahedral) — and the corresponding coordination numbers, i.e., L:2, TP:3, SP:4, O:6, etc. To account for the overall structural anisotropy, we further categorized the polyhedral connectivity[32]–from 0-dimension (0D) to 3D–based on how these polyhedra are connected within the crystal (**Fig. 2b, c**).

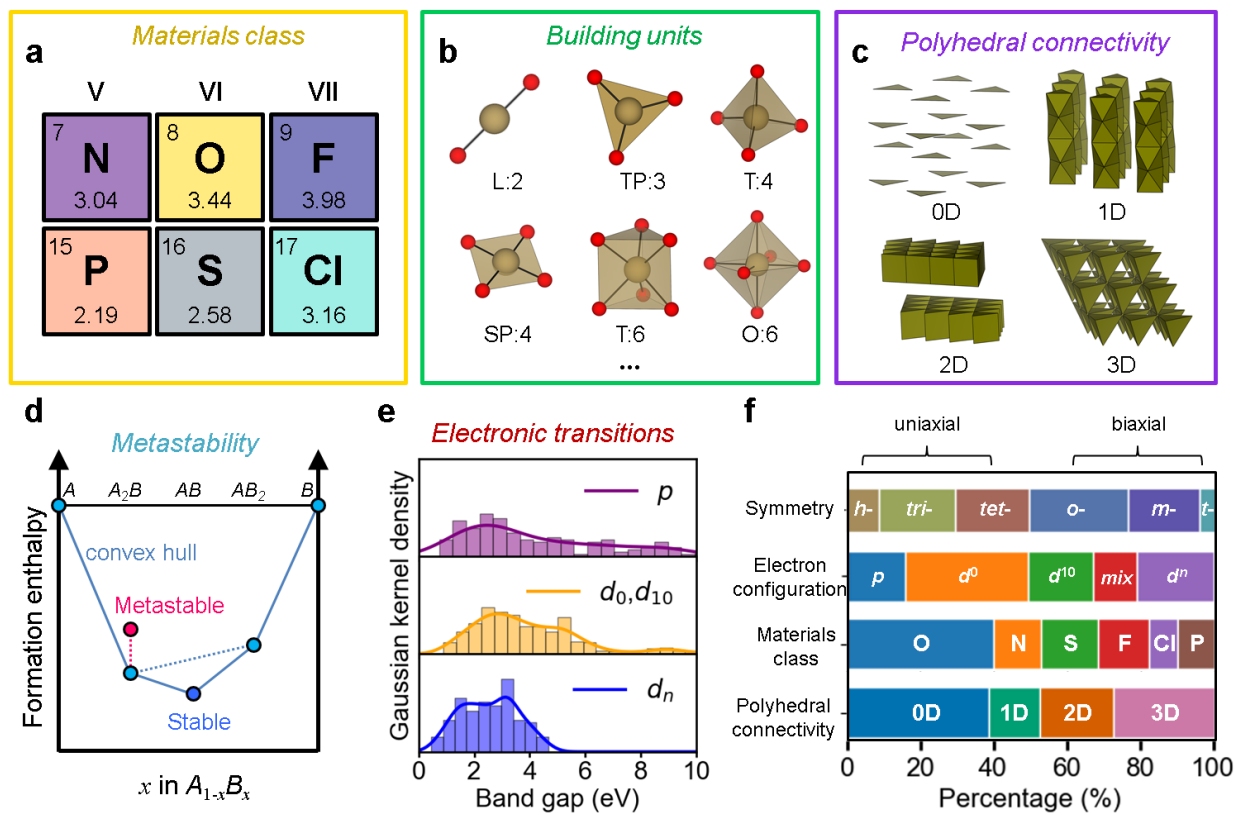

**Figure 2**. Different criteria used to categorize known birefringent compounds. (a) Electronegativity of the anions used in this study. (b) Building units as labeled with local coordination geometry of the polyhedra (linear, trigonal planar, tetragonal, square planar, trigonal prismatic, octahedral) and their corresponding coordination number (L:2, TP:3, T:4, SP:4, T:6, O:6). (c) Representative structural frameworks with different polyhedral connectivity. (d) Schematic showing a formable compound evaluated by convex hull

analysis. (e) Histograms of compounds based on the character of the valence band fitted with Gaussian kernel density distributions. (f) Statistical summary of model systems considered in this study. Crystal symmetry, electronic configuration, materials class and polyhedral connectivity are considered as design criteria.

We focus on those non-cubic crystals that have either been already synthesized or are predicted to be formable. The formability limit refers to the maximum energy above the convex hull below which a compound can be expected to exist without decomposing into other phases, as shown schematically in **Fig. 2d**. We data-mined compounds from the Materials Project database[25] that are within the formability limit specific to each material class, as reported by Sun et al.,[24] i.e., 50 meV/atom for oxides, 60 meV/atom for sulfides, 40 meV/atom for phosphides, 50 meV/atom for chlorides, 100 meV/atom for fluorides, and 200 meV/atom for nitrides. We excluded any compounds from Materials Project that have a theoretical band gap below 0.2 eV as they may show metallic behavior and lead to poor optical transparency. We also excluded heteroanion compounds to narrow down our scope to the six material classes discussed above. After applying these criteria, we filtered out 950 crystals, and we added 17 additional compounds by substituting $A$- and/or $M$-site cations in specific structural families (i.e., $A'_3MN_3$, $AMN_2$, $AMO_2$, and $A'MO_2$, where $A$ = Li, Na, K; $A'$ = Ca, Sr, Ba; $M$ = V, Nb, Ta, Ag, Au), leading to a total of 967 compounds. Of these, more than two-thirds (657 crystals) have been experimentally reported. We further classify these compounds based on their crystallographic symmetry into uniaxial and biaxial crystals. Uniaxial crystals include trigonal, hexagonal, and tetragonal space groups, and possess a single optic axis, while biaxial crystals, include orthorhombic, monoclinic, triclinic space groups, and have two optic axes.

As discussed in **Section 2.1**, the magnitude of $\Delta n$ is predominantly determined by the anisotropic distribution of the topmost valence electrons.[17, 18] In contrast, dichroism arises from the difference in absorbance along two different orientations of a crystal. It is therefore related to optical transitions from the occupied states to the empty states. DFT with semi-local exchange correlation functionals, such as the PBE functional that we have used for computational efficiency, is well known to underestimate the band gap. To obtain more accurate band gap estimates, without using the computationally expensive hybrid functionals, we employed a machine learning (ML) model developed by Wang et al.[33]. This model extrapolates band gaps calculated using the PBE functional[34] to the level of Heyd-Scuseria-Ernzerhof (HSE06) hybrid functional that gives better estimates of the band gap.[35] As shown in **Fig. S3**, there is an approximately linear relationship between the predicted HSE06 band gaps and the calculated PBE band gaps, described by the regression equation $E_g^{HSE} = 1.16 E_g^{PBE} + 0.86$, with a coefficient of determination ($R^2$) of 0.96 and a root mean square error of 0.28 eV. We note that these corrected band gaps still represent the electronic band-to-band transition, and do not account for any phonon-assisted absorption[36] or excitonic absorption near the band edge[37]. Therefore, the cutoff wavelengths derived from these predicted band gaps should be regarded as upper limits for the actual transparency window.

An analysis of the electronic structure of the screened compounds shows that the band gap, and consequently, the optical transparency, is intricately related to the electronic configuration. We classified the 967 compounds into (i) *p*-compounds where only *s*-, *p*-electron states are involved in the optical transitions, i.e., the valence and conduction bands are both made up of primarily *s*- or *p*-electronic states; (ii) $d^0$ and $d^{10}$ compounds having either empty or fully filled *d*-states at one of the band edges; (iii) *mixed* compounds having a combination of $d^0$ and $d^{10}$ transition metal (TM) cations; and (iv) $d^n$ compounds ($0 < n < 10$) having partially filled *d*- states. In *p*-compounds, the optical transitions are typically *p-p* or *p-s* transitions. In $d^0$ compounds, the associated transitions are from anion *p*-states to empty TM *d*-states. In $d^{10}$ compounds, TM *d*-states are typically deep in energy and do not participate directly in transitions near the band gap. In such compounds, the transitions involve either *p-p* or *p-s* transitions. Most $d^n$ compounds involve TM *d*-states at both the valence and conduction bands, and involve *d-d* transitions. A histogram of the corrected band gaps for the four groups is shown in **Figure 2e**. We have a fitted a Gaussian kernel density to illustrate the distribution of the typical band gap for the four groups, i.e., how the electronic configuration influences the band gap and the position of the optical absorption edge. They help us to choose materials with optical transparency for specific wavelength regions. Evidently, Mott insulators with partially filled $d^n$ configuration tend to have a narrower distribution of band gaps, primarily spanning from IR to visible regions. In contrast, band insulators with *p*-states and fully empty or filled *d*-states, which undergo *p-p, p-s* or *p-d* transitions, typically involve a broader distribution of band gaps, and are more likely to include birefringent compounds in the UV region.

We next investigated the relationship between $\Delta n$ and the predicted band gaps for the 967 compounds classified into the four groups based on their electronic configuration (**Fig. S4**). Compounds with a mix of $d^0$ and $d^{10}$ electron configuration (colored in green in **Fig. S4**), which involve *p-p, p-s* or *p-d* transitions, have a relative smaller $\Delta n$. In contrast, compounds with $d^n$ configuration (colored in blue) having *d-d* electronic transitions and narrow band gaps tend to exhibit large $\Delta n$. Such compounds have an anisotropic distribution of valence electrons, which is driven by the crystal-field splitting imposed by the coordination geometry around the TM cation. When the crystal-field induces preferential occupation of specific *d*-states (e.g., $d_{z^2}$, $d_{xy}$, or $d_{x^2-y^2}$), the electron density becomes directionally oriented, enhancing the dielectric response along specific crystallographic axes. This directional dependence can be further enhanced when the constituent polyhedra are arranged in an anisotropic manner. By comparison, traditional strategies for enhancing $\Delta n$ have focused on inducing polarizable $p$-electron states, such as those found in stereochemically active lone pairs[21] or functional anions such as carbonates[19] or borates, and typically result in modest values of $\Delta n < 0.3$. The directional nature of $d$-electronic states provides an additional avenue for achieving high $\Delta n$, particularly for visible and IR applications.

In summary, we have identified electron configuration, materials class, and polyhedral connectivity as factors that can lead to large $\Delta n$. In the following section (**Section 2.3**), we analyze how the 967 compounds are distributed across these factors and extract trends with $\Delta n$.

**2.3. Trends in optical anisotropy with changes in the key factors.** $\Delta n$ is the anisotropy in a material's dielectric response. To estimate $\Delta n$, we focused on the high-frequency dielectric tensor components ($\varepsilon^\infty$), obtained from dielectric function calculations (see the Methods section for details). The refractive indices ($n_i$) along principal axes are related to the dielectric tensor as: $n_i = \sqrt{\varepsilon_{ii}^\infty}$, where $i = x, y, z$. $\Delta n$ is then expressed as: $\Delta n = \sqrt{\varepsilon_{max}^\infty} - \sqrt{\varepsilon_{min}^\infty}$, where $\varepsilon_{min}^\infty$ and $\varepsilon_{max}^\infty$ are, respectively, the minimum and maximum value of the high-frequency dielectric tensor along different principal axes of a crystal. To quantify the anisotropy, we defined the dielectric anisotropy ($\eta$) as $\eta = 1 - \varepsilon_{min}^\infty/\varepsilon_{max}^\infty$. A value of $\eta = 0$ indicates that the material's dielectric response is fully isotropic whereas $\eta \to 1$ means that the dielectric response is highly anisotropic, as shown in **Fig. S5.**

We next analyzed how $\eta$ of the 967 compounds, which were either available on the Materials Project database or calculated, varied with the three key factors: polyhedral connectivity, materials class, and electron configuration (**Fig. 3**). In our analysis, we defined low anisotropy as $\eta < 0.3$, moderate anisotropy as $0.3 \leq \eta < 0.6$, and high anisotropy as $\eta \geq 0.6$. Firstly, with regard to polyhedral connectivity, we observe that structures with reduced dimensionality, such as quasi-1D or quasi-2D materials, exhibit large $\eta$, as shown in **Fig. 3a**. In contrast, quasi-0D and quasi-3D structures tend to be more isotropic, resulting in smaller values of $\eta$. Secondly, when considering the materials class, we find that compounds with relatively low electronegativity of the constituent – such as nitrides, sulfides, and phosphides – typically exhibit higher $\eta$, compared to those with high electronegativity, such as fluorides, oxides, and chlorides. This trend suggests that chemical systems with more covalent bonding characters, which have not been extensively explored as oxide systems, may be promising candidates for achieving large $\Delta n$. Lastly, we find that the electron configuration significantly affects $\eta$. Compounds without transition metals (*p*-compounds) and those with fully-filled *d*-states ($d^{10}$ compounds) exhibit moderate $\eta$. Compounds with empty *d*-states ($d^0$ compounds) have lower $\eta$. Compounds with mixed $d^0$ and $d^{10}$ character show low to moderate $\eta$. We find that $d^n$ compounds distinctly show large $\eta$ because their *d*-states lead to narrower gaps. In these systems, the small band gap ($E_g$) results in a high refractive index (*n*), as captured by the empirical Moss relation ($n^4 \propto 1/E_g$).[38] When this high *n* is combined with an anisotropic distribution of the valence electron states, materials with lower band gap tend to have larger $\eta$, as shown in **Fig. 3b**. We also note that the overall trends observed for $\eta$ are mirrored in $\Delta n$ (**Fig. S6**).

To further investigate the variation within the $d^n$ compounds, which was shown to exhibit large $\eta$ due to narrower band gaps, we subdivided the compounds according to their *d*-electron configuration and analyzed the distribution in their $\Delta n$ (**Fig. S7**). Among the partially filled cases, compounds with $d^2$, $d^6$ and $d^8$ count appear more frequently in our dataset, while the remaining compounds with other *d*-electron counts are grouped into a residual category labeled $d^x$. These partially filled $d^n$ compounds generally exhibit a broader $\Delta n$ distribution and include a larger portion of highly-birefringent crystals ( $\Delta n > 0.3$), with $d^2$ compounds showing notable concentrations. In contrast, compounds with $d^0$, $d^{10}$, and the mixed $d^0/d^{10}$ counts showed narrower $\Delta n$ distribution, typically below 0.3. This contrasting trend likely arises from the directional nature of partially filled *d*-states, which, under crystal-field splitting, can lead to anisotropic electron distributions and enhanced optical anisotropy.

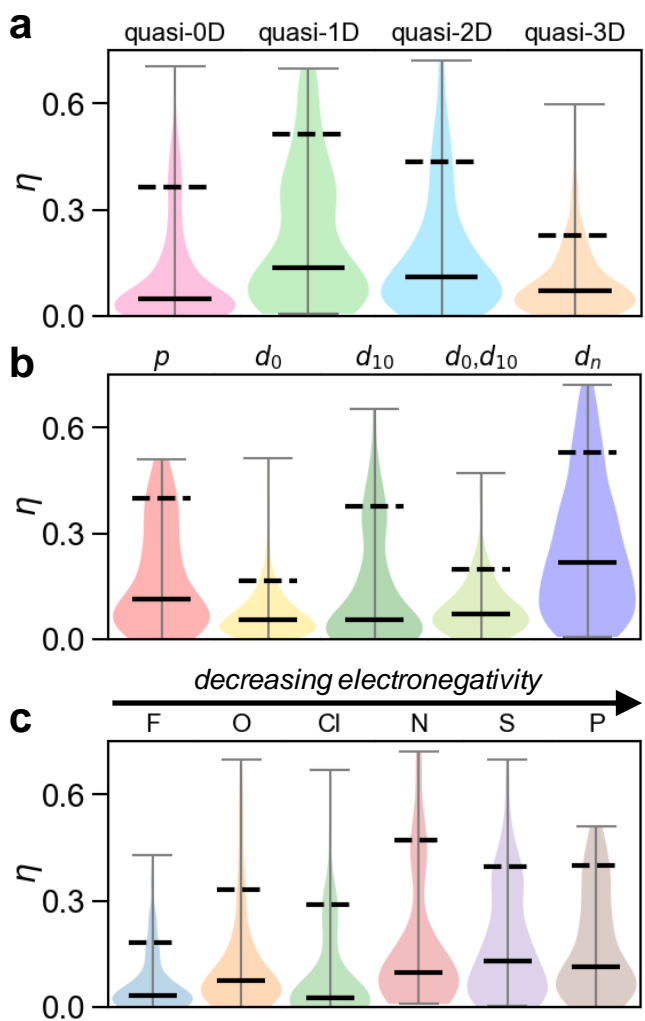

**Figure 3**. Overall trends of dielectric anisotropy ($\eta$) with three key factors. Each violin plot shows the full distribution of $\eta$ within that category. The black solid and dotted line indicates the median and the 90th

percentile of the distribution, respectively. (a) Trends by polyhedral connectivity (0D, 1D, 2D, and 3D), (b) Trends by electron filling, where the '*p*' refers to non-TM compounds, and '$d_0, d_{10}$' refers to TM compounds with mixed $d_0$ and $d_{10}$ TM cations. (c) Trends by materials class, including fluorides (F), oxides (O), chloride (Cl), nitrides (N), sulfides (S), and phosphides (P), in decreasing order of electronegativity.

**2.4. Representative birefringent compounds at different wavelengths.** To identify birefringent materials with transparency for different spectral regimes, we present a scatter plot of $\Delta n$ *versus* band gap for all the 967 compounds (**Figure 4a** and **Fig. S4a**). The overall trend line shows an inverse relation between band gap and $\Delta n$. This observation aligns with the empirical Moss relation ($n^4 \propto 1/E_g$)[38], which suggests that materials with smaller band gaps tend to exhibit higher refractive indices. To further investigate this correlation, we compared different fitting models in **Fig. S8.** Compounds with smaller band gap tend to exhibit larger $\Delta n$. This can be attributed to the presence of Mott insulators, many of which are $d^n$ compounds with partially filled *d*-states. When these compounds experience crystal-field splitting, they are likely to have more directional and anisotropic electron distributions, as discussed in Section 2.3. We have highlighted promising compounds within each spectrum (UV, visible and IR), and listed them in **Tables S1-S3**. These compounds are categorized by their local coordination environment using the labelling system discussed previously. In the IR region, highly birefringent compounds are found among $d^n$ compounds, especially those having anisotropic polyhedral arrangements such as quasi-1D or quasi-2D compounds. Specific examples exhibiting $\Delta n \geq 1.0$ include $Sr_xTiS_3$ ($x$ = 9/8, 8/7 and 6/5), for which all three phases lie on the convex hull and exhibit band gaps ranging from 1.3 to 1.5 eV (cutoffs: 830–950 nm); $ARuN$ ($A$ = Na, Li), which are 0.086 – 0.152 eV/atom above the hull with band gaps of ~1.2 eV (cutoff: ~1030 nm); and $A'_3MN_3$ ($A'$= Ca, Sr; $M$ = V, Nb, Ta), with the energy above the hull ($E_{\text{hull}}$) ranging from 0 to 0.059 eV/atom and band gaps between 1.2 and 1.7 eV (cutoffs: 730–1030 nm). A density-of-states (DOS) analysis reveals that both $Sr_xTiS_3$ and $A'_3MN_3$ ($A'$ = Ca, Sr) possess filled $d_{z^2}$ non-bonding states near the Fermi energy (**Fig. S9a,b**). These $d_{z^2}$ states form a highly oriented electron blob along the optic axis, contributing to strong anisotropic polarizability and a large $\Delta n$. Meanwhile, $A$RuN ($A$ = Na, Li) compounds with a $d^6$ configuration exhibit selective occupation of in-plane $d$-states ($d_{xz}$, $d_{yz}$, $d_{xy}$, $d_{x^2-y^2}$) near the Fermi energy (**Fig. S9c**). This directional occupation enhances the dielectric response along the $a$-$b$ plane and leads to the large $\Delta n$.

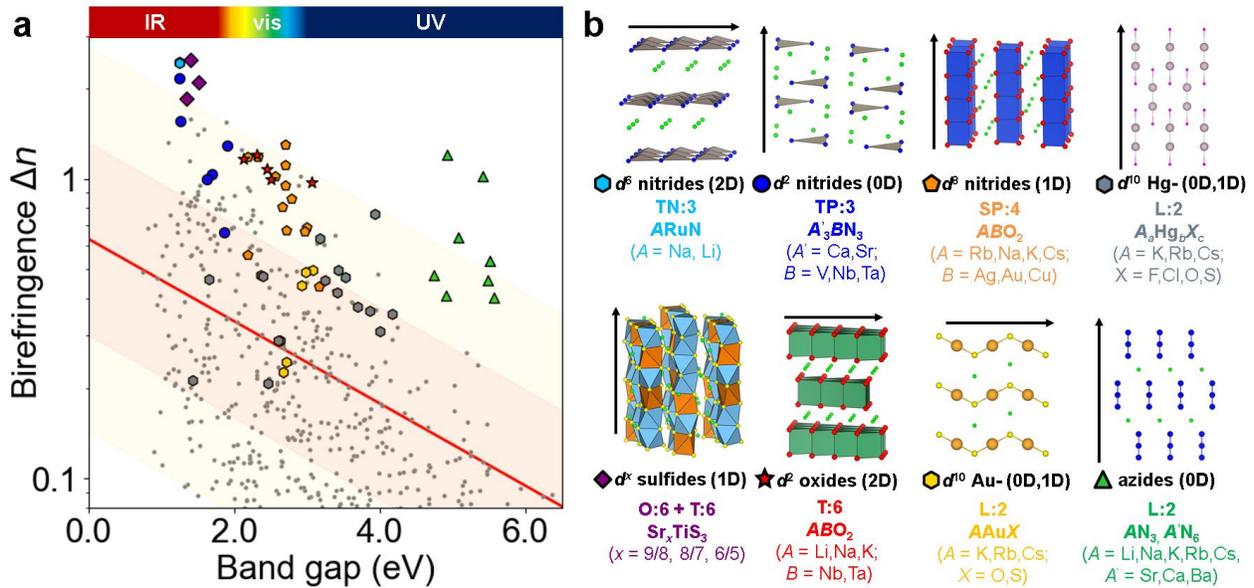

**Figure 4**. (a) Birefringence ($\Delta n$) *versus* predicted band gap of 967 compounds. Red trend line shows a linear fit to all data points, and the shaded regions represent the $\pm 1\sigma$ (darker shading) and $\pm 2\sigma$ (lighter shading) intervals of the fit. (b) Representative compounds with large $\Delta n$ for IR, vis, and UV regions. The local coordination environment (local geometry and coordination number), materials class, $d$-electron numbers, and polyhedral connectivity of the representative compounds are shown in (b). Black arrows represent the axis of highest refractive index. Color schemes - oxides: red; nitrides: blue; sulfides: yellow; halides: pink; alkaline or alkaline earth metals: green.

In the visible region, compounds such as $AMO_2$ ($A$ = Li, Na, K; $M$ = Nb, Ta) with $d^2$ configuration having $E_{hull}$ = 0 – 0.005 eV/atom and band gaps from 2.3 to 3.1 eV (cutoffs: 400–540 nm); and $AMO_2$ ($A$ = Rb, Na, K, Cs; $M$ = Ag, Au, Cu) with $d^8$ configuration, where all are on the hull with band gaps ranging from 2.2 to 2.7 eV (cutoffs: 460–560 nm), show $\Delta n \geq 0.6$. The former has 2D layered connectivity, while the latter exhibit quasi-1D connectivity through square-planar building units. DOS analysis indicates that the anisotropy in these compounds arises from the selectively occupied $d_{z^2}$ or $p_z$ states near the Fermi energy. Due to the structural arrangement of the polyhedra, these occupied electronic states are oriented within the *a-b* plane (**Fig. S9d,e**), and amplify the material's dielectric response along this plane, resulting in the large $\Delta n$.

In the UV region, Au- and Hg-containing compounds having a $d^{10}$ configuration were identified exhibiting band gaps between 3 and 4 eV (cutoffs: 310–410 nm) and $\Delta n$ between 0.2 to 0.8, with the $E_{hull}$ within 0.012 eV/atom. Additionally, a family of azides ($AN_3$ and $A'N_6$, where $A$ = Li, Na, K, Rb, Cs and $A'$ = Sr, Ca) were identified, where all of these compounds are on the hull and have band gaps > 4 eV (cutoff < 310 nm) and $\Delta n \geq 0.4$. The Au- and Hg-containing $d^{10}$ compounds and the azides possess linear coordination geometries. We attribute the large $\Delta n$ of the identified azides to the splitting between the in-plane $p_x, p_y$ states — that are degenerate (forming $\pi$-type molecular orbitals)— and the out-of-plane $p_z$ states (forming $\sigma$-type molecular orbitals, see **Fig. S9h**)[39]. This orbital splitting leads to an anisotropic electron distribution that enhances $\Delta n$. A similar splitting of the *p*-orbitals is also observed in the Au- and Hg-containing compounds (**Fig. S9g**), but they are less anisotropic due to the mixed contributions from *d*-electron states near the Fermi energy.

Across all spectral regions, a commonality of the highly birefringent compounds is their anisotropic, low-dimensional crystal structures (quasi-1D or quasi-2D) and the presence of selectively filled electronic states along specific crystallographic directions near the Fermi energy. This selective occupation is promoted by the crystal-field splitting. Thus, a potential approach to achieve giant optical anisotropy is to engineer the anisotropic arrangement of electronic states by modulating both the spatial arrangement and electron filling of the polyhedral units within the crystal structure, which we discuss more detail in the following section.

**2.5. Strategies to achieve giant optical anisotropy.** In this section, we discuss design strategies to achieve giant optical anisotropy. To do this, we compared select birefringent compounds spanning the IR, visible and UV regions – $Ca_3TaN_3$, $LiNbO_2$, and $NaN_3$, respectively, – and compared them to isostructural compounds that exhibit less- or negligible $\Delta n$ ($Ca_3GaN_3$, $LiGaO_2$, and $CsN_3$). The crystal structure of these six compounds is shown in **Fig. 5a-c**. As previously discussed, electron filling influences the distribution of the polarizable electrons that contribute to the dielectric response, and impacts $\Delta n$. Highly birefringent compounds, such as $Ca_3TaN_3$ ($\Delta n = 2.17$) and $LiNbO_2$ ($\Delta n = 1.17$), both of which have a $d^2$ configuration, exhibit non-bonding $d_{z^2}$ states at the top of their valence bands (left panels in **Fig. 5a,b**). These states are oriented along specific crystallographic directions: along the *c*-axis in $Ca_3TaN_3$ and within the *a-b* plane in $LiNbO_2$. Conversely, their isostructural counterparts, $Ca_3GaN_3$ ($\Delta n = 0.4$) and $LiGaO_2$ ($\Delta n = 0.001$), both having a $d^{10}$ configuration, showed much reduced $\Delta n$. These reductions arise because the valence bands near the Fermi energy are dominated by more isotropic *p*-electron states (right panels in **Fig. 5a,b**), as we discussed in **Section 2.1**. In $Ca_3GaN_3$, the N. $2p_x$ and N. $2p_y$ states are evenly distributed in the *a–b* plane, leading to the reduced anisotropy. In $LiGaO_2$, the *p*- electron character is even more isotropic along all the three axes, resulting in a nearly negligible Δ$n$.

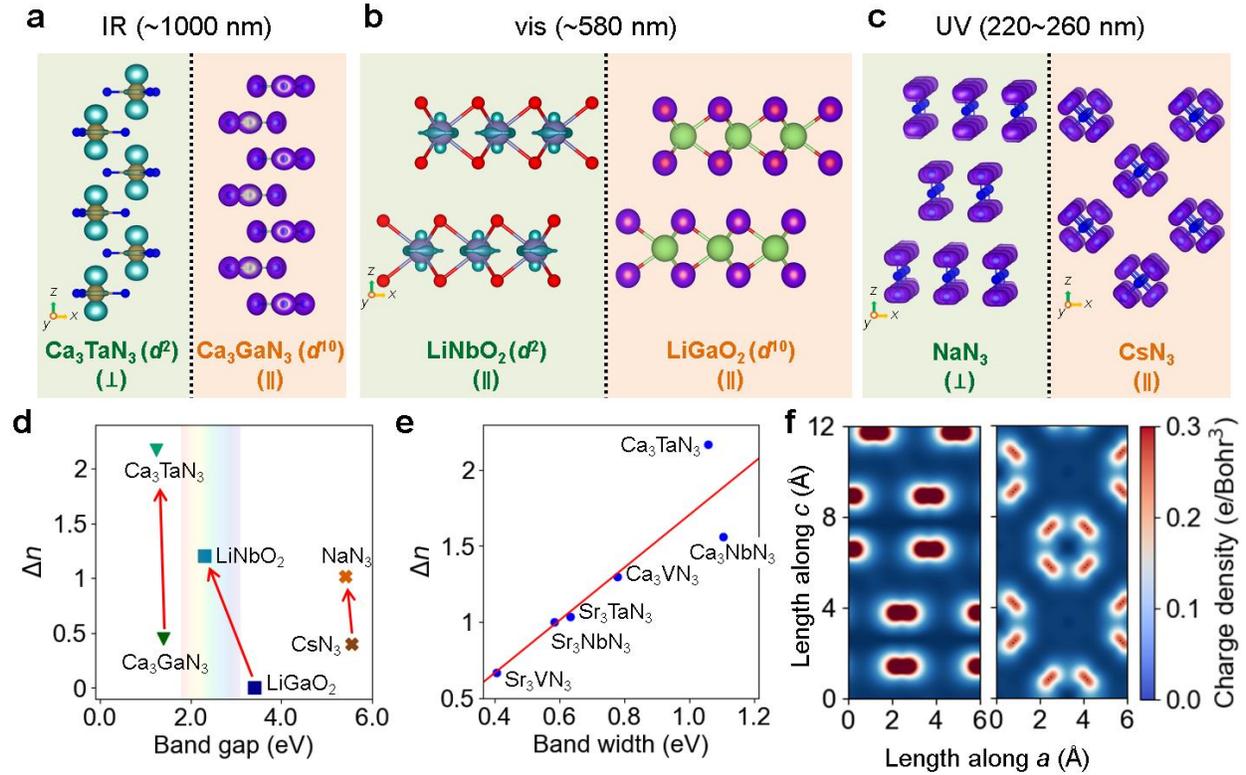

Figure 5. (a-c) Atomic structures of select birefringent compounds on the left and their less- or non-birefringent counterparts on the rights for three different spectrum: (a) $Ca_3TaN_3$ and $Ca_3GaN_3$ in the IR spectrum; (b) $LiNbO_2$ and $LiGaO_2$ in the visible spectrum; (c) $NaN_3$ and $CsN_3$ in the UV spectrum. Spatial distribution of the valence electrons below the Fermi energy are visualized by isosurfaces. (d) **Δ$n$** *versus* the bandgap of selected compounds. (e) **Δ$n$** *versus* bandwidths along the out-of-plane direction (Γ-Z) for **$A'_3MN_3$** compounds ($A'$ = Ca, Sr; $M$ = V, Nb, Ta). (f) 2D-projected charge density isosurface of azides ($NaN_3$ and $CsN_3$).

Another strategy to achieve large Δ$n$ is to modulate the anisotropy of the electronic band structure in the reciprocal space. As an example, we analyzed the electronic band structure of a series of $A'_3MN_3$ compounds ($A'$= Ca, Sr; $M$ = V, Nb, Ta), each having a $d^2$ configuration of the

Page 20 of 28

TM ions (**Fig. S10**). We quantified the dispersion of the topmost valence band along the high symmetry line, corresponding to the out-of-plane direction Γ-Z. This dispersion is captured by measuring the bandwidth, which correlates with $\Delta n$, as shown in **Fig. 5e**. For example, $Ca_3TaN_3$ exhibits a large band dispersion with a width of 1.06 eV and the contribution near the Fermi energy solely comes from the out-of-plane (Γ-Z) direction (**Fig. S10**), leading to the large $\Delta n = 2.2$. In comparison, $Sr_3VN_3$ shows a reduced anisotropy in band dispersion with a width of 0.40 eV along the out-of-plane direction (Γ-Z), with significant contribution from the electronic bands along the in-plane directions (Γ-S-Y-Γ) near the Fermi energy. These factors lead to a smaller optical anisotropy, with a reduced $\Delta n = 0.66$. These results demonstrate how tuning the anisotropy of band dispersion through careful selection of chemical composition can optimize the optical anisotropy.

Lastly, the spatial density and the arrangement of the electron clouds close to the Fermi energy also affect $\Delta n$. To illustrate this point, we use select compounds from the azide family ($AN_3$, $A$ = Li, Na, K, and Cs) as examples. In these compounds, the arrangement of $N_3^-$ anions is determined by the *A*-site cations. Smaller cations like $Li^+$ and $Na^+$ favor a linear arrangement of the $N_3^-$ anions, while larger cations like $Cs^+$ and $K^+$ induce zigzag geometries due to steric effects. We plotted the 2D-projected charge density of all of the filled valence states (which represents the electron density distribution in a plane) along the *a-c* plane (**Fig. 5f** and **Fig. S11**). Highly birefringent compounds, such as $LiN_3$ and $NaN_3$, exhibit linear arrangements of $N_3^-$, with pronounced charge densities along the optic axis (*c*-axis), resulting in large $\Delta n$ (**Fig. S11a,b**). In contrast, compounds like $CsN_3$ and $KN_3$, with zigzag geometries, show a more isotropic charge distribution, leading to a reduced $\Delta n$ (**Fig. S11c,d**). These findings suggest that modulating the spatial arrangement of the anions by changing the *A*-site cations could be a general strategy to enhance optical anisotropy in other functional anionic systems such as carbonates[19] and borates[20], where high $\Delta n$ has also been reported.

## 3. CONCLUSIONS

In this study, we systematically investigated 967 compounds to identify highly birefringent materials across the UV to IR spectrum. More than two-thirds of these compounds are experimentally synthesizable and all of them lie within the formability limit. Among these compounds, 216 potential birefringent crystals are newly identified with $\Delta n > 0.3$. By combining high-throughput screening with electronic structure analysis, we evaluate the impact of electronic configuration, polyhedral connectivity, and chemical composition on the optical anisotropy. Our result show that the anisotropic distribution of topmost electrons determines the $\Delta n$, and it is maximized by modulating the electron filling and optimizing the spatial arrangement of those polarizable electron states. These findings provide a basis for the rational design of highly birefringent materials, including the identified compounds such as $AMO_2$, $A'MN_3$ ($A$ = Li, Na, K; $A'$ = Ca, Sr, Ba, and $M$ = Nb, Ta) and azide family. Future research could focus on experimental synthesis and characterization of the predicted highly-birefringent compounds to validate our results. Since the distilled design principles are based on general features of electronic states near the Fermi energy and their spatial arrangement, these principles are not limited to specific materials class or structural motifs. Thus, the same design strategies can be extended to more complex anion systems and organic-inorganic hybrid compounds, which further enrich the pool of materials with giant optical anisotropy. Particularly, the strategies described here could be also applied to discover birefringent polymers for flexible and lightweight optical applications. Ultimately, integrating these materials into optical devices could advance applications in photonics and laser technologies.

## 4. METHODS

**4.1. DFT calculations.** All the calculations were performed using projector augmented-wave potentials as implemented in the plane-wave DFT code, VASP[40, 41]. The PBE functionals[34] within the generalized gradient approximation (GGA) was used to describe the exchange-correlation interactions. A plane-wave basis with an energy cutoff of 600 eV was used with an electronic convergence criteria of $10^{-7}$ eV. Structures were relaxed until all the forces on the atoms were less than 0.01 eV/Å. $k$-points spacing of 0.04 and 0.025 Å$^{-1}$ were chosen for the structural relaxations and static calculations, respectively. An effective on-site Hubbard $U$ approach[42] (with the rotationally invariant scheme) was used for the following transition metals: Co (3.32 eV), Cr (3.7 eV), Fe (5.3 eV), Mn (3.9 eV), Ni (6.2 eV), and V (3.25 eV). We calculated the birefringence as the difference between ordinary ($n_o$) and extraordinary ($n_e$) indices ($\Delta n = n_e - n_o$). Assuming an independent particle picture, the frequency-dependent dielectric function was computed along different crystallographic axes using the LOPTICS tag in VASP, which uses the formulation proposed by Gajdoš et al.[43] To ensure computational efficiency and accuracy, we selected an energy spectrum for optical transitions by setting the total number of energy bands (NBANDS) to be 2.5 times the number of valence bands. We evaluated the density of states using 8000 grid points along the energy axis (NEDOS tag in VASP) and employed a complex shift (CSHIFT tag in VASP) value of 0.1 for the Kramers-Kronig transformation. To evaluate the accuracy of our calculated $\Delta n$ values, we compared them with available experimental data for several representative crystals, including calcite, Sr$_{9/8}$TiS$_3$, α-BaB$_2$O$_4$, AgGaS$_2$, and rutile TiO$_2$, as shown in **Fig. S12**. The calculated $\Delta n$ values show excellent agreement with experimental results, with deviations within ~0.07.

**4.2. Band gap correction.** To improve the accuracy of the band gap predictions beyond the limitations of the PBE functional, we employed the ML model developed by Wang *et al*.[33] This model was trained on the SNUMAT dataset,[44] which includes the band gaps of 10,481 inorganic compounds calculated at the HSE06 level, covering unary, binary, ternary, quartenary, quinary, and higher-order materials. Wang *et al*.[33] used a random forest regression algorithm, and the model validation was performed using 5-fold cross-validation. The input features used for training the ML model included the calculated band gaps at the PBE level and elemental properties such as electronegativity, atomic mass, melting point, atomic radius, and ionization energy.

## ASSOCIATED CONTENT

**Supporting Information**

The Supporting Information is available free of charge on the ACS Publications website.

Orbital-projected band structures for prototype structures; coordination geometries and crystal field splitting; band gap corrections via machine learning; birefringence versus dielectric anisotropy trends; overall trends of birefringence with key factors; fitting models for the birefringence versus band gap; electronic band structure and DOS analysis; charge density isosurface; experimental versus theoretical birefringence; tables of birefringent compounds in UV, visible, and IR region.

## AUTHOR INFORMATION

**Corresponding Author**

Email: gjung@wustl.edu (G.Y.J), rmishra@wustl.edu (R.M.)

**Author Contributions**

G.Y.J. and R. M. conceived the idea and initiated the project. G.R. and P. O. contributed to data analysis and interpretation. J. R. contributed to the discussion of results. G.Y.J. and R.M. drafted the manuscript with edits from all authors.

**Notes**

The authors declare no conflict of interest.

**Data availability**

The data that support the findings of this study are openly available at https://10.5281/zenodo.15161709, reference number.[45]

**ACKNOWLEDGEMENTS**

This work was primarily supported the National Science Foundation through grant nos. DMR-2122070 (G.Y.J., G.R., R.M.), DMR-2122071 (J.R.), and DMR-2145797 (P.O., R.M.). This work used computational resources through allocation DMR160007 from the Advanced Cyberinfrastructure Coordination Ecosystem: Services & Support (ACCESS) program, which is supported by NSF grants #2138259, #2138286, #2138307, #2137603, and #2138296.

**Table of contents (TOC) graphic**

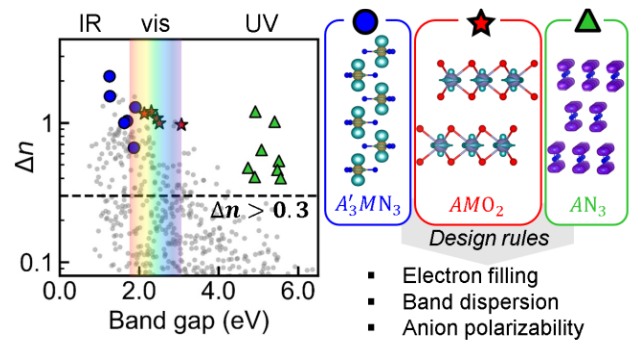